\documentclass{INTERSPEECH2023}
\usepackage{multirow}
\interspeechcameraready

\title{Experimenting with Additive Margins for Contrastive\\ Self-Supervised Speaker Verification}
\name{Theo Lepage, Reda Dehak}
\address{
  Speaker and Language Recognition Group (ESLR), \\
  Laboratoire de Recherche de l'EPITA, France
}

\email{\{theo.lepage, reda.dehak\}@epita.fr}

\begin{document}

\maketitle
 
\begin{abstract}
Most state-of-the-art self-supervised speaker verification systems rely on a contrastive-based objective function to learn speaker representations from unlabeled speech data. We explore different ways to improve the performance of these methods by: (1) revisiting how positive and negative pairs are sampled through a “symmetric” formulation of the contrastive loss; (2) introducing margins similar to AM-Softmax and AAM-Softmax that have been widely adopted in the supervised setting. We demonstrate the effectiveness of the symmetric contrastive loss which provides more supervision for the self-supervised task. Moreover, we show that Additive Margin and Additive Angular Margin allow reducing the overall number of false negatives and false positives by improving speaker separability. Finally, by combining both techniques and training a larger model we achieve 7.50\% EER and 0.5804 minDCF on the VoxCeleb1 test set, which outperforms other contrastive self supervised methods on speaker verification.\\
\end{abstract}

\section{Introduction}

\newcommand\blfootnote[1]{%
  \begingroup
  \renewcommand\thefootnote{}\footnote{#1}%
  \addtocounter{footnote}{-1}%
  \endgroup
}

\blfootnote{The code associated with this article is publicly available at\\ \url{https://github.com/theolepage/sslsv}.}

Speaker Recognition (SR) aims to recognize the identity of the person speaking on an input speech audio. It is a fundamental task of speech processing and finds its wide applications in real-world voice-based authentication of persons. Different speech feature extraction methods and machine learning frameworks were proposed for this task. Learning speaker embedding space \cite{villalbaStateoftheartSpeakerRecognition2020} is the trend of speaker recognition, which has been widely developed in several aspects. The i-vectors \cite{dehakLanguageRecognitionIvectors2011}, the d-vector \cite{varianiDeepNeuralNetworks2014}, and the x-vectors \cite{snyderXVectorsRobustDNN2018, chungDelvingVoxCelebEnvironment2020, chungDefenceMetricLearning2020} were proposed to represent the speaker variability. The i-vector is a generative method trained in an unsupervised manner. The other approaches discriminatively embed speakers into a vector space using deep neural networks that require large labeled datasets. Although impressive progress has been made with supervised learning, this paradigm is now considered a bottleneck for building more intelligent systems. Manually annotating data is complex, expensive, and tedious, especially when dealing with signals such as images, text, and speech. Moreover, the risk is creating biased models that do not work well in real life, notably in difficult acoustic conditions.

Recently, motivated by the surge of self-supervised learning concepts, many deep embedding methods \cite{stafylakis19_interspeech, Cho2020, xiaSelfSupervisedTextIndependentSpeaker2021, zhangContrastiveSelfSupervisedLearning2021, choNonContrastiveSelfsupervisedLearning2022} have proven to be very effective in benefiting from the massive amount of unlabeled data. Like classical approaches, most self-supervised learning methods aim to learn an embedding space that maximizes the similarity between embeddings of similar inputs and minimizes the similarity between embedding of different inputs without human supervision. When dealing with audio samples, the assumption is that segments extracted from the same utterance belong to the same speaker, but those from different utterances belong to distinct speakers. This assumption does not always hold (\textit{class collision} issue), but the impact on the training convergence is negligible. Segments extracted from the same utterance share different information, such as channel, language, speaker, and sentiment information \cite{9893562}. Speech augmentation is necessary in this case to help the algorithm ignore channel characteristics and focus only on speaker-related information.

In Speaker Verification (SV), different self-supervised methods have been proposed. Methods based on a contrastive loss, such as SimCLR \cite{zhangContrastiveSelfSupervisedLearning2021}, MoCo \cite{xiaSelfSupervisedTextIndependentSpeaker2021} and VICReg \cite{lepageLabelEfficientSelfSupervisedSpeaker2022}, have been successfully applied to this field of research. These methods are based on the Normalized Temperature-scaled Cross Entropy (NT-Xent) loss, making distances between positive pairs small and between negative pairs large in a latent space. The majority of these approaches have focused on how to define the model architecture and sample negative pairs for the training. 

Different objective functions were proposed in supervised speaker recognition, and an effort has been made to improve softmax-based classification losses to learn better representations. Angular-based losses have become popular and compute the cosine similarity by normalizing embedding vectors and the output layer. Inspired by face recognition, angular margin-based losses have also been successfully applied to supervised speaker recognition \cite{utterancelevelaggforsv, largemarginsoftmaxforsv, 8683649} to improve the angular softmax loss. Introducing a margin in the angular softmax loss achieves promising results when selecting an appropriate margin scale, as it increases the separability between speakers.

In this paper, we propose to introduce Additive Margin and Additive Angular Margin into the SimCLR training framework \cite{zhangContrastiveSelfSupervisedLearning2021}. We adopt the NT-Xent loss used in the literature and define SNT-Xent-AM and SNT-Xent-AAM to experiment with varying values of margin. Moreover, we show that using a “symmetric” formulation of the contrastive loss, by using all possible positive and negative pairs, improves the downstream performance. Our training framework is further described in Section~\ref{sec:method}. In Section~\ref{sec:ExpSetup}, we present our experimental setup. We report our results and assess the effect of margins in Section~\ref{sec:ResDisc}. Furthermore, we show that we can achieve competitive results compared to state-of-the-art contrastive methods while using a simple framework and relying only on the VoxCeleb1 \cite{nagraniVoxCelebLargeScaleSpeaker2017} dev set. Finally, we conclude in Section~\ref{sec:conclusion}.

\section{Method}
\label{sec:method}

\begin{figure}[ht]
  \centering
  \includegraphics[width=\linewidth]{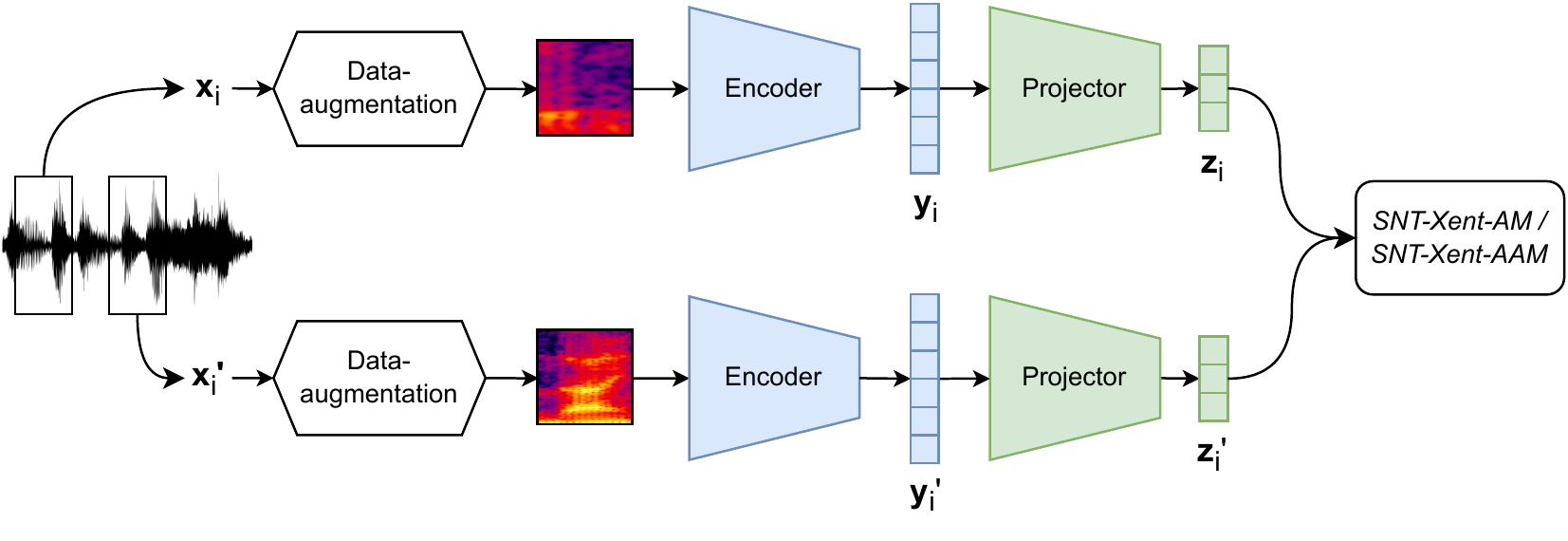}
  \caption{Diagram of our contrastive self-supervised training framework to learn speaker representations.}
  \label{fig:training_framework}
\end{figure}

Our self-supervised training framework is depicted in Figure~\ref{fig:training_framework}. The architecture uses a simple siamese neural network to produce a pair of embeddings for a given unlabeled utterance.

For each training step, we randomly sample $N$ utterances from the dataset. Let $i \in I \equiv\{1 \ldots N\}$ be the index of a random sample from the mini-batch.  We extract two non-overlapping frames, denoted as $\boldsymbol{x}_i$ and $\boldsymbol{x}_i^{\prime}$, from each utterance.
Then, we apply random augmentations to both copies and use their mel-scaled spectrogram as features for the neural network. Using different frames and applying data augmentation is fundamental to avoid collapse and to produce channel invariant representations that only depend on speaker identity.
An encoder transforms $\boldsymbol{x}_i$ and $\boldsymbol{x}_i^{\prime}$ to their respective representations $\boldsymbol{y}_i$ and $\boldsymbol{y}_i^{\prime}$. Then, the representations are fed to a projector to compute the embeddings $\boldsymbol{z}_i$ and $\boldsymbol{z}_i^{\prime}$. During training, mini-batches are created by stacking $\boldsymbol{z}_i$ samples into $\boldsymbol{Z}$ and $\boldsymbol{z}_i^{\prime}$ samples into $\boldsymbol{Z'}$.

Representations are used to perform speaker verification, while the embeddings are used to calculate the loss and optimize the model.

\subsection{Contrastive-based self-supervised learning}
\label{subsec:CSSL}

Contrastive learning aims at maximizing the similarity within positive pairs while maximizing the distance between negative pairs. In self-supervised learning, supervision is provided by assuming that each utterance in the mini-batch belongs to a unique speaker. Positive pairs are constructed with embeddings derived from the same utterances, while negative pairs are sampled from other elements in the mini-batch.

We start by defining the similarity between two embeddings $\boldsymbol{u}$ and $\boldsymbol{v}$ as $\ell(\boldsymbol{u}, \boldsymbol{v})=e^{\cos\left(\theta_{\boldsymbol{u}, \boldsymbol{v}}\right) / \tau}$ where $\theta_{\cdot, \cdot}$ is the angle between two vectors and $\tau$ is a temperature scaling hyper-parameter. $\cos(\theta_{\cdot, \cdot})$ corresponds to the cosine similarity and is obtained by computing the dot product between two $l_2$ normalized embeddings.

Then, the Normalized Temperature-scaled Cross Entropy loss (NT-Xent) is defined as
\begin{equation}
    \mathcal{L}_{\text {NT-Xent}}=- \frac{1}{N} \sum_{i \in I} \log \frac{\ell \left(\boldsymbol{z}_i, \boldsymbol{z}_{i}^{\prime} \right)}{\sum_{a \in I} \ell \left(\boldsymbol{z}_i, \boldsymbol{z}_{a}^{\prime} \right)} \label{eq:eqn1}
\end{equation}

We refer to $\boldsymbol{z}_i$ as the \textit{anchor}, $\boldsymbol{z}_i^{\prime}$ as the \textit{positive}, and $\boldsymbol{z}_a^{\prime}$ as the \textit{negatives}. Thus, a total of $N$ positive pairs are created, and each is compared to $N-1$ negatives.

\subsection{Maximizing the supervisory signal provided to the self-supervised learning task}
\label{subsec:SNTXent}

The previous formulation of the contrastive loss, used in \cite{zhangContrastiveSelfSupervisedLearning2021}, does not consider all possible positive and negative pairs. Following SimCLR \cite{chenSimpleFrameworkContrastive2020}, we used the “symmetric” formulation of the NT-Xent loss to increase the number of comparisons and maximize the supervisory signal provided to the self-supervised objective function.
 
We now consider $\boldsymbol{z}_i$ to be the $i$-th element of a set of all embeddings created by concatenating $\boldsymbol{Z}$ and $\boldsymbol{Z}^{\prime}$. Let $i \in \hat{I} \equiv\{1 \ldots 2 N\}$ be the index of a random augmented sample, $j(i)$ be the index of the other augmented sample originating from the same mini-batch and $A(i) \equiv \hat{I} \setminus \{i\}$. The SNT-Xent loss is defined as
\begin{equation}
    \mathcal{L}_{\text {SNT-Xent}}=- \frac{1}{2N} \sum_{i \in \hat{I}} \log \frac{\ell \left(\boldsymbol{z}_i, \boldsymbol{z}_{j(i)}\right)}{\sum_{a \in A(i)} \ell \left(\boldsymbol{z}_i, \boldsymbol{z}_a \right)} \label{eq:eqn2}
\end{equation}

Note that this framework generates $2N$ positives pairs (each utterance and its other augmented version), and each one of them is compared to a set of $2(N-1)$ negatives (the other utterances except the positive and the anchor). This is interesting as having more contrastive examples has been shown to produce a better performance on the downstream task.

Finally, we propose to compute the similarity of positive pairs and negative pairs differently, making it easier to introduce future improvements (\emph{Margin} in the next section) to the objective function such that
\begin{multline}
\footnotesize
\mathcal{L}_{\text {SNT-Xent}}=- \frac{1}{2N}\\ \sum_{i \in \hat{I}} \log \frac{\ell^{+} \left(\boldsymbol{z}_i, \boldsymbol{z}_{j(i)}\right)}{\ell^{+} \left(\boldsymbol{z}_{i} , \boldsymbol{z}_{j(i)} \right) + \sum_{a \in \hat{A}(i)} \ell^{-} \left(\boldsymbol{z}_{i} , \boldsymbol{z}_a \right)}\label{eq:eqn3}
\end{multline}
where $\ell^{+}(\mathbf{u}, \mathbf{v}) = \ell^{-}(\mathbf{u}, \mathbf{v}) = e^{\cos \left(\theta_{\boldsymbol{u}, \boldsymbol{v}}\right) / \tau}$ and $\hat{A}(i) \equiv \hat{I} \setminus \{i, j(i)\}$.

This loss and its variants are at the core of all self-supervised contrastive learning frameworks. However, it aims to penalize classification errors instead of producing discriminative representations which would be relevant in the context of speaker verification.

\subsection{Introducing margins to improve speaker separability}
\label{subsec:Margin}
We explore two ways to improve the discriminative capacity of a contrastive-based objective function using the SNT-Xent loss as our baseline. Inspired by state-of-the-art techniques for face recognition, we introduce margins to increase the similarity of same-speaker embeddings further. These methods were successfully applied for training end-to-end speaker verification models in a supervised way \cite{utterancelevelaggforsv, largemarginsoftmaxforsv, 8683649} which justifies our motivation to adapt these concepts for self-supervised learning.
\subsubsection{Additive Margin}
\label{subsubsec:am}
Following CosFace \cite{cosface}, we introduce an extra margin in \textit{cosine space} to force the cosine similarity of positive pairs to be above a specific threshold and thus improve speaker separability. 

To illustrate the effect of this technique, we consider a scenario using the non-symmetric version of NT-Xent and setting $N=2$. Therefore, in this example, we consider a total of 2 classes based on the self-supervised contrastive assumption. In the case of the first speaker, the NT-Xent loss forces $\cos \left(\theta_{\boldsymbol{z}_1, \boldsymbol{z}_1^{\prime}}\right) > \cos \left(\theta_{\boldsymbol{z}_1, \boldsymbol{z}_2^{\prime}}\right)$. By introducing a margin, we further require $\cos \left(\theta_{\boldsymbol{z}_1, \boldsymbol{z}_1^{\prime}}\right) - m > \cos \left(\theta_{\boldsymbol{z}_1, \boldsymbol{z}_2^{\prime}}\right)$ where $m \geq 0$ is a fixed scalar introduced to control the magnitude of the cosine margin. Intuitively, this could help the contrastive objective since the constraint is more stringent, as well as improve downstream performance by maximizing inter-speaker distance and eventually minimizing intra-speaker variance.

We refer to this loss as $\mathcal{L}_{\text{SNT-Xent-AM}}$ which is identical to SNT-Xent except that we set  $\ell^{+}(\mathbf{u}, \mathbf{v})=e^{\left(\cos\left(\theta_{\mathbf{u}, \mathbf{v}}\right)-m\right) / \tau}$ while $\ell^{-}(\mathbf{u}, \mathbf{v})$ remains unchanged.

\subsubsection{Additive Angular Margin}
Inspired by ArcFace \cite{arcface}, the second method is referred to as additive angular margin and consists in introducing the margin directly in \textit{angle space}. As opposed to the previous technique, it provides the exact correspondence to the geodesic distance.

Following the case scenario presented in the previous section, the angular margin will translate to a decision boundary for the first speaker of the form $\cos \left(\theta_{\boldsymbol{z}_1, \boldsymbol{z}_1^{\prime}} + m \right) > \cos \left(\theta_{\boldsymbol{z}_1, \boldsymbol{z}_2^{\prime}}\right)$, where $m \geq 0$ is a fixed scalar introduced to control the magnitude of the angular margin.

To train the model with additive angular margin we rely on the loss $\mathcal{L}_{\text{SNT-Xent-AAM}}$ which is identical to SNT-Xent except that we use $\ell^{+}(\mathbf{u}, \mathbf{v})=e^{\cos \left(\theta_{\mathbf{u}, \mathbf{v}} + m \right) / \tau}$ and keep  $\ell^{-}(\mathbf{u}, \mathbf{v})$ unchanged.

We observed training instability when optimizing SNT-Xent-AAM from random initialization, especially with large values of $m$. We hypothesize that reducing the difficulty of the self-supervised task early in the training is fundamental for allowing the loss to converge. Thus, we gradually increase the margin for our experiments from 0 to its final value during the first half of the training with a cosine scheduler. This strategy could be referred to as a kind of curriculum learning, and similar techniques have already been employed to solve this issue.

\section[Experimental setup]{Experimental setup}
\label{sec:ExpSetup}
\subsection{Datasets and feature extraction}
\label{subsec:datafeat}
Considering the training time\footnote{Limited by our computing power, we had to restrict our experiments to the VoxCeleb1 training set.}, we train our model on the VoxCeleb1 dev set, which contains $148,642$ utterances from $1,211$ speakers. The evaluation is performed on the VoxCeleb1 ‘original’ test set composed of $4,874$ utterances from $40$ speakers. Speaker labels are discarded during self-supervised training. From audio chunks of $2$ seconds per sample, we extract $40$-dimensional log-mel spectrogram features with a Hamming window of $25 ms$ length with a $10 ms$ frame-shift. We do not apply Voice Activity Detection (VAD) as training data consists mostly of continuous speech segments. The network input features are normalized using instance normalization.

\subsection{Data-augmentation}
\label{subsec:dataaug}
To produce representations robust against extrinsic variabilities, self-supervised learning frameworks commonly rely on extensive data-augmentation techniques. In the context of speaker verification, we aim to learn embeddings invariant to channel information, such as noise from the environment or recording device. Therefore, providing different views of the same utterance is crucial to avoid encoding channel characteristics, allowing speaker identity to be the only distinguishing factor between two representations. During training, we randomly apply a range of transformations to the input signal at each training step. We add background noise, overlapping music tracks, or speech segments using the MUSAN corpus. To simulate various real-world scenarios to augment the utterances, we randomly sample the Signal-to-Noise Ratio (SNR) between $\left[13; 20\right]$ dB for speech, $\left[5; 15\right]$ dB for music, and $\left[0; 15\right]$ dB for noises. To further enhance our self-supervised model’s robustness, we apply reverberation to the augmented utterances using the simulated Room Impulse Response database.

\subsection{Models architecture and training}
\label{subsec:model}
First, to run more experiments, we used a Thin ResNet-34 architecture with $512$ output units for the encoder. We rely on self-attentive pooling (SAP) to generate utterance-level representations. The projector consists of a standard MLP, composed of two fully-connected layers with $2048$ and $256$ units, respectively. The intermediate layer is followed by ReLU nonlinearity. We optimize the model using the Adam optimizer with a learning rate of $0.001$, which is reduced by $5\%$ every $10$ epochs, with no weight decay. We use a batch size of $256$ and train the model for $200$ epochs. Our implementation is based on the PyTorch framework, and we conduct our experiments using 2x NVIDIA Titan X (Pascal) $12 GB$. Regarding the loss computation, we use $\tau=0.02$ as the temperature hyper-parameter by default. For the final results, we train for $300$ epochs a larger ResNet-34 model using channel dimension blocks twice as large as the smaller encoder.

\subsection{Evaluation protocol}
\label{subsec:evalprotocol}
To evaluate our model's performance on speaker verification, we extract embeddings from a fixed number of evenly spaced frames for each test utterance before averaging them across the temporal axis. Then, we compute the cosine similarity between two $l_2$-normalized embeddings to determine the scoring. Following VoxCeleb and NIST Speaker Recognition evaluation protocols, we report the performance of our model in terms of Equal Error Rate (EER) and minimum Detection Cost Function (minDCF) with $P_{target} = 0.01$, $C_{\text{miss}}=1$ and $C_{\text{fa}}=1$.

\section{Results and discussions}
\label{sec:ResDisc}

\subsection{The effect of the different self-supervised training components}
\label{subsec:ResSNT-Xent}

We conduct an ablation study to assess the role of the different components of our self-supervised training framework and report the results in Table~\ref{tab:SNT-Xent}. The NT-Xent loss, which is our baseline, achieves $9.45\%$ EER and $0.7094$ minDCF. First, we verify that applying data-augmentation is fundamental for learning relevant representations and that training the model with a projector produces better performance. Then, we show that relying on more positive pairs and negative pairs with the symmetric contrastive loss results in $9.35\%$ EER and $0.6647$ minDCF. This validates our intuition that providing more supervision is beneficial to improve the self-supervised system's downstream performance. This system will be used as the baseline for the next experiments.

\begin{table}[th]
  \caption{The effect of data-augmentation, projecting representations during training, and the symmetric formulation of the contrastive objective function on speaker verification results (Thin ResNet-34 encoder).}
  \label{tab:SNT-Xent}
  \centering
  \begin{tabular}{lcc}
    \toprule
    \textbf{Method}     &     \textbf{EER(\%)}     & \textbf{minDCF} \\
    \midrule
    Baseline    &  $9.45$  & $0.7094$ \\
    Baseline w/o Data-augmentation   &  $28.17$ & $0.8656$ \\
    Baseline w/o Projector           &  $13.55$ & $0.8435$ \\
    Baseline w/ SNT-Xent & $\boldsymbol{9.35}$ & $\boldsymbol{0.6647}$ \\
    \bottomrule
  \end{tabular}
\end{table}

\subsection{Results on speaker verification when introducing margins in the self-supervised contrastive loss}
\label{subsec:ResMargins}
The choice of the margin value has a significant impact on speaker verification results. As shown in Table~\ref{tab:ResMargins}, the best setting is $m=0.4$ for Additive Margin and $m=0.1$ for Additive Angular Margin, achieving $8.70\%$ EER and $8.98\%$ EER, respectively. For both methods, a small margin has no effect on the results but a very large margin prevents learning good speaker representations. It is noteworthy that this does not correspond to the default value often used for supervised training which is $m=0.2$. In particular, the Additive Angular Margin is more sensitive to the margin factor and suffers from exploding gradients with a margin greater than or equal to $0.3$. This result is understandable as the margin is applied in \textit{angle space}. Note that learning the margin value jointly with the model degrades the performance. Finally, we observe that introducing margins reduces the EER, resulting in fewer false positives and false negatives overall. However, this improvement does not translate on the minDCF. Note that standard DNN-based speaker embedding extractors are not optimized to improve the minDCF during training. Thus, margins can be incorporated into the self-supervised contrastive training to improve results on speaker verification. We hypothesize that other downstream tasks related to verification could benefit from this method.

\begin{table}[th]
  \caption{Speaker verification results when introducing margins in the self-supervised contrastive loss (Thin ResNet-34 encoder).}
  \label{tab:ResMargins}
  \centering
  \begin{tabular}{lccc}
    \toprule
    \textbf{Loss} & \textbf{Margin} & \textbf{EER(\%)} & \textbf{minDCF} \\
    \midrule
    SNT-Xent                       &   -   &  $9.35$ & $0.6647$ \\
    \midrule
    \multirow{6}{*}{SNT-Xent-AM}   & $0.1$ & $9.30$  & $0.7610$ \\
    ~                              & $0.2$ & $9.01$  & $0.6907$ \\
    ~                              & $0.3$ & $8.93$  & $0.6909$ \\
    ~                              & $0.4$ & $\boldsymbol{8.70}$ & $\boldsymbol{0.6873}$ \\
    ~                              & $0.5$ & $8.87$ & $0.7182$ \\
    ~                              & \textit{\footnotesize Learnable} & $9.26$ & $0.7093$ \\
    \midrule
    \multirow{5}{*}{SNT-Xent-AAM} & $0.05$ & $8.92$ & $0.7006$ \\
    ~                             & $0.1$ & $\boldsymbol{8.98}$ & $\boldsymbol{0.6742}$ \\ 
    ~                             & $0.2$ & $9.22$ & $0.6846$ \\
    ~                             & $0.3$ & \multicolumn{2}{c}{\textit{\footnotesize Exploding gradients}} \\
    ~                             & \textit{\footnotesize Learnable} & $9.18$ & $0.6717$ \\
    \bottomrule
    \end{tabular}
\end{table}

\subsection{Study of scores distribution}
\label{subsubsec:ScoreDist}

\begin{figure}[t]
  \centering
  \includegraphics[width=0.85\linewidth]{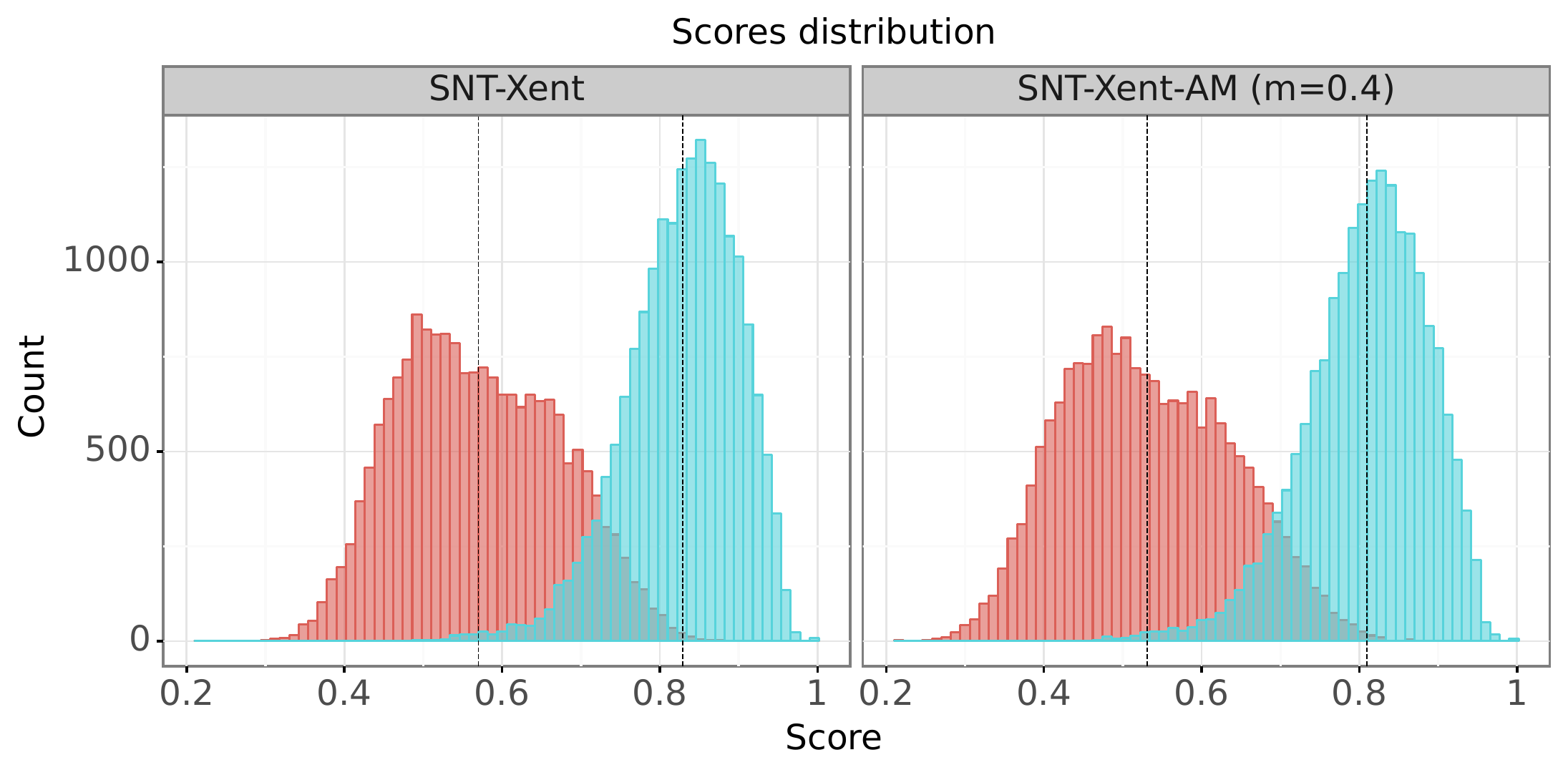}
  \caption{Positive (light blue) and negative (red) trials scores distribution obtained after training with SNT-Xent and SNT-Xent-AM ($m=0.4$) losses. The mean of each distribution is represented by a vertical dashed line.}
  \label{fig:AMScoreDist}
\end{figure}

In Figure~\ref{fig:AMScoreDist}, we plot the distribution of scores computed on the test set to assess the effect of margins on the learned representations. Visually, we can notice that the spread between the distribution of positive and negative scores is further when using Additive Margin with a margin of $0.4$. This result was expected since our method aims at separating positive from negative scores. The difference between the mean of the positives and the mean of the negatives trials scores is $0.259$ without margins (SNT-Xent) while it reaches $0.278$ with margins (SNT-Xent-AM). This is consistent with the improvement of the EER and shows that margins have an effect on the discriminative power of self-supervised systems designed for verification.

\subsection{Comparison to other self-supervised contrastive methods for speaker verification}
We report the final results on speaker verification in Table~\ref{tab:cmpSSL}. By training for more epochs and using a larger encoder, we reach $7.56\%$ EER with the symmetric contrastive loss (SNT-Xent) and with additive angular margin (SNT-Xent-AAM) while we achieve $7.50\%$ EER with additive margin (SNT-Xent-AM). This corresponds to a $13.8\%$ relative improvement of the EER compared to the model trained with SNT-Xent-AM during our early experiments. Furthermore, our method outperforms other works based on contrastive learning for self-supervised speaker verification while using a smaller training set, i.e., VoxCeleb1. This result implies that standard contrastive methods can be further improved by introducing several changes designed explicitly for self-supervised learning (symmetric contrastive loss) and speaker verification (additive margin).

\begin{table}[th]
   \caption{Comparison of self-supervised contrastive methods for speaker verification. Our methods are trained on VoxCeleb1 while the first three systems were trained on VoxCeleb2 ($\sim7\times$ more samples).}
   \label{tab:cmpSSL}
   \centering

   \begin{tabular}{lcc}
    \toprule
    \textbf{Method} & \textbf{EER(\%)}     & \textbf{minDCF} \\
     \midrule
    
      AP+AAT \cite{huhAugmentationAdversarialTraining2020}   &  $8.65$ & $-$ \\
    SimCLR \cite{zhangContrastiveSelfSupervisedLearning2021} & $8.28$ & $0.6100$ \\
    MoCo \cite{xiaSelfSupervisedTextIndependentSpeaker2021} & $8.23$ & $0.5900$ \\
    
    \midrule
    SNT-Xent & $7.56$ & $\boldsymbol{0.5785}$ \\
    SNT-Xent-AM ($m=0.4$) & $\boldsymbol{7.50}$ & $0.5804$\\
    SNT-Xent-AAM ($m=0.01$) & $7.56$ & $0.6281$ \\
     \bottomrule
   \end{tabular}
\end{table}

\section{Conclusion}
\label{sec:conclusion}
In this paper, we proposed an improvement of self-supervised contrastive frameworks to learn more robust speaker representations. First, we demonstrated that providing more self-supervision with additional positive and negative pairs through the SNT-Xent loss is essential to get better performances. Next, we showed that introducing margins in the contrastive loss function leads to a lower EER on the VoxCeleb1 test dataset and a better discrepancy between scores of positive and negative trials. The performance of our larger final model with additive margin is competitive with other self-supervised contrastive techniques for speaker verification.

\bibliographystyle{IEEEtran}
\bibliography{mybib}

\end{document}